\documentclass[Physsubmission, Phys]{SciPost}
\hypersetup{
    colorlinks,
    linkcolor={red!50!black},
    citecolor={blue!50!black},
    urlcolor={blue!80!black}
}

\usepackage[bitstream-charter]{mathdesign}
\newcommand{\ii}{\mathrm i}
\newcommand{\ee}{\mathrm e}

\DeclareSymbolFont{usualmathcal}{OMS}{cmsy}{m}{n}
\DeclareSymbolFontAlphabet{\mathcal}{usualmathcal}

\begin{document}

\begin{center}{\Large \textbf{
Josephson effects between the Kitaev ladder superconductors\\
}}\end{center}

% TODO: write the author list here. Use initials + surname format.
% Separate subsequent authors by a comma, omit comma at the end of the list.
% Mark the corresponding author with a superscript *.
\begin{center}
Osamu Kanehira\textsuperscript{1} and
Hiroki Tsuchiura\textsuperscript{1,2$\star$}
\end{center}

\begin{center}
{\bf 1} Department of Applied Physics, Tohoku University, Sendai, Japan
\\
{\bf 2} Center for Spintronics Research Network, Sendai, Japan
\\
* tsuchi@tohoku.ac.jp
\end{center}

\begin{center}
\today
\end{center}

% For convenience during refereeing (optional),
% you can turn on line numbers by uncommenting the next line:
%\linenumbers

\definecolor{palegray}{gray}{0.95}
\begin{center}
\colorbox{palegray}{
  \begin{tabular}{rr}
  \begin{minipage}{0.1\textwidth}
    \includegraphics[width=30mm]{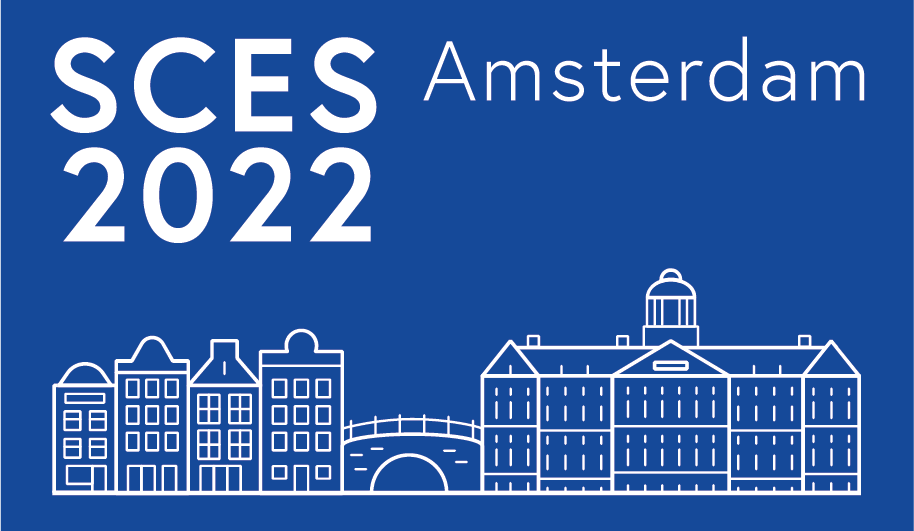}
  \end{minipage}
  &
  \begin{minipage}{0.85\textwidth}
    \begin{center}
    {\it International Conference on Strongly Correlated Electron Systems\\ (SCES 2022)}\\
    {\it Amsterdam, 24-29 July 2022} \\
    \doi{10.21468/SciPostPhysProc.?}\\
    \end{center}
  \end{minipage}
\end{tabular}
}
\end{center}

\section*{Abstract}
{\bf
The two-leg ladder system consisting of the Kitaev chains is known to exhibit a richer phase diagram than that of the single chain. 
We theoretically investigate the variety of the Josephson effects between the ladder systems.
We consider the Josephson phase difference $\theta$ between these two ladder systems as well as the phase difference $\phi$ between the parallel chains in each ladder system.
The total energy of the junction at $T = 0$ is calculated by a numerical diagonalization method as functions of $\theta$, $\phi$, and also a transverse hopping $t_{\perp}$ in the ladders.
We find that, by controlling $t_{\perp}$ and $\phi$, the junction exhibits not only the fractional Josephson effect for the phase difference $\theta$, but also the usual 0-junction and even $\pi$-junction properties. 
}

\vspace{10pt}
\noindent\rule{\textwidth}{1pt}
\tableofcontents\thispagestyle{fancy}
\noindent\rule{\textwidth}{1pt}
\vspace{10pt}

\newpage
%%%%%%%%%%%%%%%%%%%%%%%%%%%%%%%%%%
\section{Introduction}
\label{sec:intro}
%%%%%%%%%%%%%%%%%%%%%%%%%%%%%%%%%%
% TODO: write your article here.
%The stage is yours. Write your article here.
%The bulk of the paper should be clearly divided into sections with short descriptive titles, including an introduction and a conclusion.
%%%%%%%%%%%%%%%%%%%%%%%%%%%%%%%%%%%%%%%%%%%%%

A topological Josephson junction consisting of two Kitaev superconducting chains exhibits a fractional Josephson effect with a 4$\pi$ periodicity in the current-phase relationship \cite{kitaev2001}.
This is a promising experimental proof for the existence of Majorana zero modes (MZMs) at the edges of the Kitaev chains. 
Experimentally, several methods have been proposed to realize the Kitaev chains, and a promising method is to create $p$-wave pairings by the proximity effect between an $s$-wave superconducting substrate with strong spin-orbit interaction and semiconductor nanowires \cite{alicea2012}. 
Furthermore, theoretically, multi-terminal Josephson junctions of Kitaev chains have been proposed to realize the so-called braiding operation of MZMs to perform topological quantum calculations \cite{alicea2011}. 
In such hybrid systems, the Majorana fermion signature may appear as a zero-bias peak in conductance, and many experimental attempts have been made \cite{das2012,mourik2012}. 
However, no conclusive evidence for the existence of MZM, let alone braiding operations, has yet been obtained.

Despite the experimental difficulty of realization and the theoretical simplicity, the Kitaev chain is still attracting considerable interest.
One direction is to study the ladder systems consisting of multiple Kitaev chains.
It has been reported that a richer phase diagram and novel phenomena can be obtained even in minimal two-leg ladder systems\cite{wakatsuki2014,maiellaro2018,nakosai2018,Nehra2019}.
The Hamiltonian of the two-leg Kitaev ladder system in general is given as
%\end{eqnarray}
\begin{align}
\label{eq:ladder_hamiltonian}
 H = & -\mu\sum_{i=1}^{N}\sum_{j=1,2}c^\dag_{i,j}c_{i,j}
    - \sum_{i=1}^{N-1}\sum_{j=1,2} \left[t\,c^\dag_{i+1,j}c_{i,j} + \Delta_j c^{\dagger}_{i+1,j}c^{\dagger}_{i,j} + \text{h.c.}\right]
\nonumber \\
    & - \sum_{i=1}^{N} \left[
    t_{\perp} c^\dag_{i,1}c_{i,2}
    + \Delta_{\perp}c_{i,1}^{\dagger}c_{i,2}^{\dagger} + \text{h.c.}
    \right]
%  H = &-\mu\sum_{i=1}^{N}\sum_{j=1,2}c^\dag_{i,j}c_{i,j} - %\frac{t_{\perp}}{2}\sum_{i=1}^{N}\big[c^\dag_{i,1}c_{i,2} + c^\dag_{i,2}c_{i,1}
%  -\frac12 \sum_{i=1}^{N-1}\sum_{j=1,2}\Big[t\,c^\dag_{i+1,j}c_{i,j} \big]\nonumber \\
%    & -\frac12 \sum_{i=1}^{N-1}\sum_{j=1,2}\Big[t\,c^\dag_{i+1,j}c_{i,j} + \Delta_j\,c^\dag_{i+1,j}c^\dag_{i,j}+\text{h.c.}\Big],
\end{align}
where $i=1,2,\cdots,N$ is the site number of the each chain and $j=1,2$ is the chain number in the Kitaev ladder.
Also, $t$ is the nearest-neighbor transfar integral, $\mu$ is the chemical potential, $\Delta_j = \Delta\ee^{\ii \phi_j}$ is the intra-chain $p$-wave superconducting pair potential with phase $\phi_j$, and $t_{\perp}$ and $\Delta_{\perp}$ is the coupling between two chains. 
%In the following, we will introduce the superconducting phase difference $\phi$, then the superconducting phase of the two chains can be chosen as $\phi_1=0,~\phi_2=\phi$. 
%This phase difference can be realized experimentally by a bias current in the rung direction of the Kitaev ladder \cite{nakosai2018}.
As is well known, in the single Kitaev chain, the topological phase is characterized by the ${\mathbb{Z}}_{2}$ invariant, and the phase diagram is described only by $\mu/t$.
%the chemical potential $\mu$ and 
%the intra-chain hopping $t$.
In contrast in the two-leg Kitaev ladder systems, a $\mathbb{Z}$ topological invariant characterizes the topological phase, and the inter-chain parameters $t_{\perp}$ and $\Delta_{\perp}$ also affect the phase diagram.
Recently, Maiellaro {\it et al.} have shown that the system exhibits a topological phase either 
with four or two Majorana zero-energy modes \cite{maiellaro2018}. 
They also find that the topological phase survives also when the Kitaev's criterion 
$\Delta > 0$ and $|\mu| < 2t$ for the single chain is violated.

Because of the multiplicity of the Majorana zero modes and topological quantum numbers, we can naturally expect that the Josephson effect in the ladder systems may exhibit various current-phase relations.
Thus in this paper, we investigate the Josephson effects in systems consisting of {\it two} two-leg Kitaev ladders by using numerical diagonalization.
Finding a variety of Josephson current-phase relationship would open up a wider range of the potential applications of the Kitaev chains other than topological quantum computation.

This paper is organized as follows.
In section 2, we introduce the model Hamiltonian for the Josephson junction considered here. 
In section 3, we show the Josephson energy-phase relationship by using numerical diagonalization of the Hamiltonian. 
%
%The Andreev bound states as functions of the Josephson phase are also shown to discuss the obtained energy-phase relationships.
Finally, we summarize our results in section 4.

\section{Model Hamiltonian}
\begin{figure}[htbp]
  \begin{center}
    \includegraphics[height=28mm]{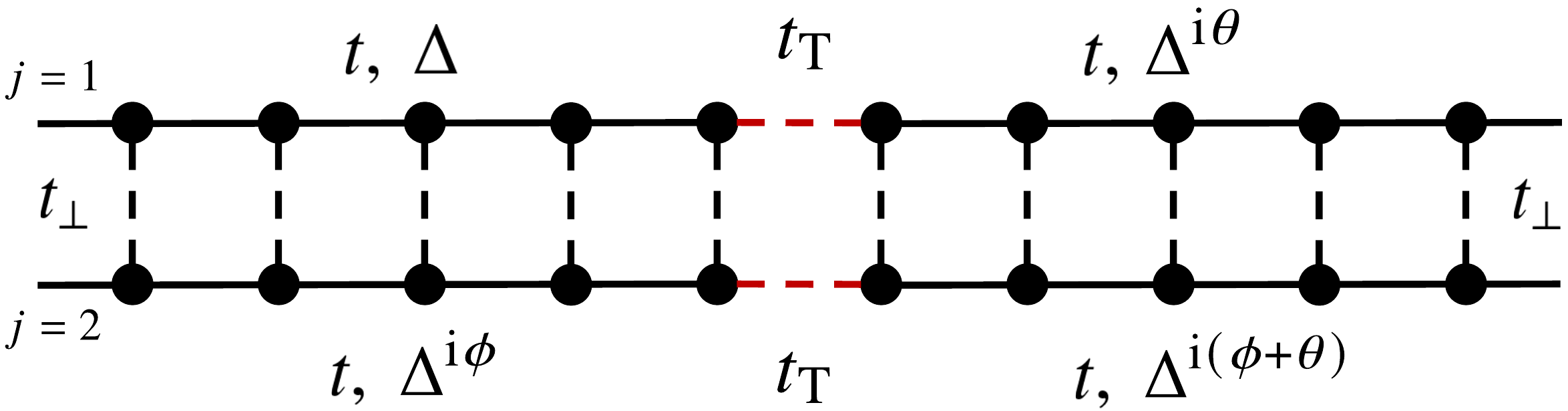}
    \caption{
    Schematic illustration of the junction system consisting of two Kitaev ladders.
    Note that the superconducting pair potential $\Delta$ is only assumed within each chain.
    Here, two superconducting phase differences are assumed; $\theta$ is the Josephson phase difference across the junction, and $\phi$ is the additional inter-chain phase difference.
    } \label{fig:junction}
  \end{center}
\end{figure}

%%%%%%%%%%%%%%%%%%%%%%%%%%%%%%%%%%%%%%%%%%%%%%
%接合系のモデル：模式図とハミルトニアン
In this section, we introduce a junction system consists of the two Kitaev ladders.
%discussed in the previous section, and calculate method of the Josephson energy $E(\theta)$, where $\theta$ is the phase difference between the Kitaev ladders. 
The Hamiltonian of the present system depicted in Fig. \ref{fig:junction} is written in the standard notation as
\begin{align}
\label{eq:total}
  &H = H_{\text L} + H_{\text R} + H_{\text T}, \\[6pt]
\label{eq:left}
  &H_{\text{L}} = -\mu\sum_{i,j}c^{\text{L}\dag}_{i,j}c^{\text{L}}_{i,j} - \frac{t_{\perp}}{2}\sum_{i}\Big[c^{\text{L}\dag}_{i,1}c^{\text{L}}_{i,2} + c^{\text{L}\dag}_{i,2}c^{\text{L}}_{i,1}\Big] \nonumber \\
    &\hspace{24pt}-\frac12 \sum_{i,j}\Big[t\,c^{\text{L}\dag}_{i+1,j}c^{\text{L}}_{i,j} + \Delta_j\,c^{\text L\dag}_{i+1,j}c^{\text L\dag}_{i,j}+\text{h.c.}\Big], \\[6pt]
\label{eq:right}
  &H_{\text{R}} = -\mu\sum_{i,j}c^{\text R\dag}_{i,j}c^{\text R}_{i,j} - \frac{t_{\perp}}{2}\sum_{i}\Big[c^{\text R\dag}_{i,1}c^{\text R}_{i,2} + c^{\text R\dag}_{i,2}c^{\text R}_{i,1}\Big] \nonumber \\
    &\hspace{24pt}-\frac12 \sum_{i,j}\Big[t\,c^{\text R\dag}_{i+1,j}c^{\text R}_{i,j} + \Delta_j\ee^{\ii\theta}\,c^{\text R\dag}_{i+1,j}c^{\text R\dag}_{i,j}+\text{h.c.}\Big], \\[6pt]
\label{eq:junction}
  &H_{\text{T}} = -\frac{t_{\text T}}{2}\sum_{j=1,2}\Big(c_{N,j}^{\text{L}\dag}c_{1,j}^{\text R} + c_{1,j}^{\text{R}\dag}c_{N,j}^{\text L}\Big),
\end{align}
where 
%$\Delta_1 = \Delta,~\Delta_2 = \Delta\ee^{\ii\phi}$ and 
$c_{i,j}^{\text L}/c_{i,j}^{\text R}$ is the annihilation operator of the electron on the $i$-th site
of the $j$-th chain to the left/right side of the junction. 
Thus the last contribution in (\ref{eq:total}), $H_{\mathrm T}$, corresponds to the tunnel Hamiltonian between the L/R ladders, given explicitly in (\ref{eq:junction}).
We note that the superconducting pair-potential $\Delta_{j}$ is only assumed within each chain.
We also assume the two superconducting phase differences $\theta$ and $\phi$; $\theta$ is the Josephson phase difference across the junction, and $\phi$ is the phase difference between the chains $j=$ 1 and 2.
It is worth noting here that this interchain phase difference $\phi$ may induce several novel phenomena
such as the modulation of the phase difference of the superconducting order parameter in two chains \cite{nakosai2018} or the increasing of crossed Andreev reflection \cite{Nehra2019}. 
Thus we introduce $\phi$ here in the hope that a novel phenomena would be induced again.
%This system is graphically described in figure \ref{fig:junction} 

%We fixed the phase difference inside each ladder to $\phi$, and the phase difference between ladders to $\theta$.

%ここで、我々は新たに鎖間の超伝導位相差\phiを導入した。前章で取り上げた先行研究ではこのような鎖間位相差は考慮されていなかったが、超伝導ペアリングの空間変調やcrossed Andreev reflectionの増大などの興味深い現象を引き起こす。\cite{nakosai2018,Nehra2019}そのため、このような鎖間位相差を接合モデルにも導入することにより、新奇な位相特性が見られることが期待される。

Throughout this paper, we take $t = 2\Delta =
1$ and $t_{\mathrm T} = 0.1$.
The total number of sites per each chain is set to $N=100$ for calculation of the Josephson effect. 
We compute the ground state energy of the whole system as a function of $\theta$ for several 
$t_{\perp}$, $\mu$, and also $\phi$ by using numerical diagonalization.

%$H_{\text L}$ and $H_{\text{R}}$ are described as 
%\begin{align}
%  H_{\text{L}} = &-\mu\sum_{i,j}c^{\text{L}\dag}_{i,j}c^{\text{L}}_{i,j} - \frac{t_{\perp}}{2}\sum_{i}\Big[c^{\text{L}\dag}_{i,1}c^{\text{L}}_{i,2} + c^{\text{L}\dag}_{i,2}c^{\text{L}}_{i,1}\Big] \nonumber \\
%   &-\frac12 \sum_{i,j}\Big[t\,c^{\text{L}\dag}_{i+1,j}c^{\text{L}}_{i,j} + \Delta\,c^\dag_{i+1,j}c^\dag_{i,j}+\text{h.c.}\Big], \\
%      H_{\text{R}} = &-\mu\sum_{i,j}c^\dag_{i,j}c_{i,j} - \frac{t_{\perp}}{2}\sum_{i}\Big[c^\dag_{i,1}c_{i,2} + c^\dag_{i,2}c_{i,1}\Big] \nonumber \\
%        &-\frac12 \sum_{i,j}\Big[t\,c^\dag_{i+1,j}c_{i,j} + \Delta\ee^{\ii\theta}\,c^\dag_{i+1,j}c^\dag_{i,j}+\text{h.c.}\Big],
%\end{align}

% numerical calculations

%We evaluated energy-phase relation of the junction system (\ref{eq:total}) by the energy difference of the system between with and without the junction $E_J(\theta) = E(\theta,t_{\text T})-E(\theta,0)$, calculated by numerical exact diagonalization.\par

%%%%%%%%%%%%%%%%%%

%%%%%%%%%%%%%%%%%%%%%%%%%%%%%%%%%%%%%%%%%%%%%%
\section{Results}
%%%%%%%%%%%%%%%%%%%%%%%%%%%%%%%%%%%%%%%%%%%%%%
%接合エネルギーと各φにおけるパターンの相図

\begin{figure}[htbp]
  \begin{center}
    \includegraphics[height=100mm]{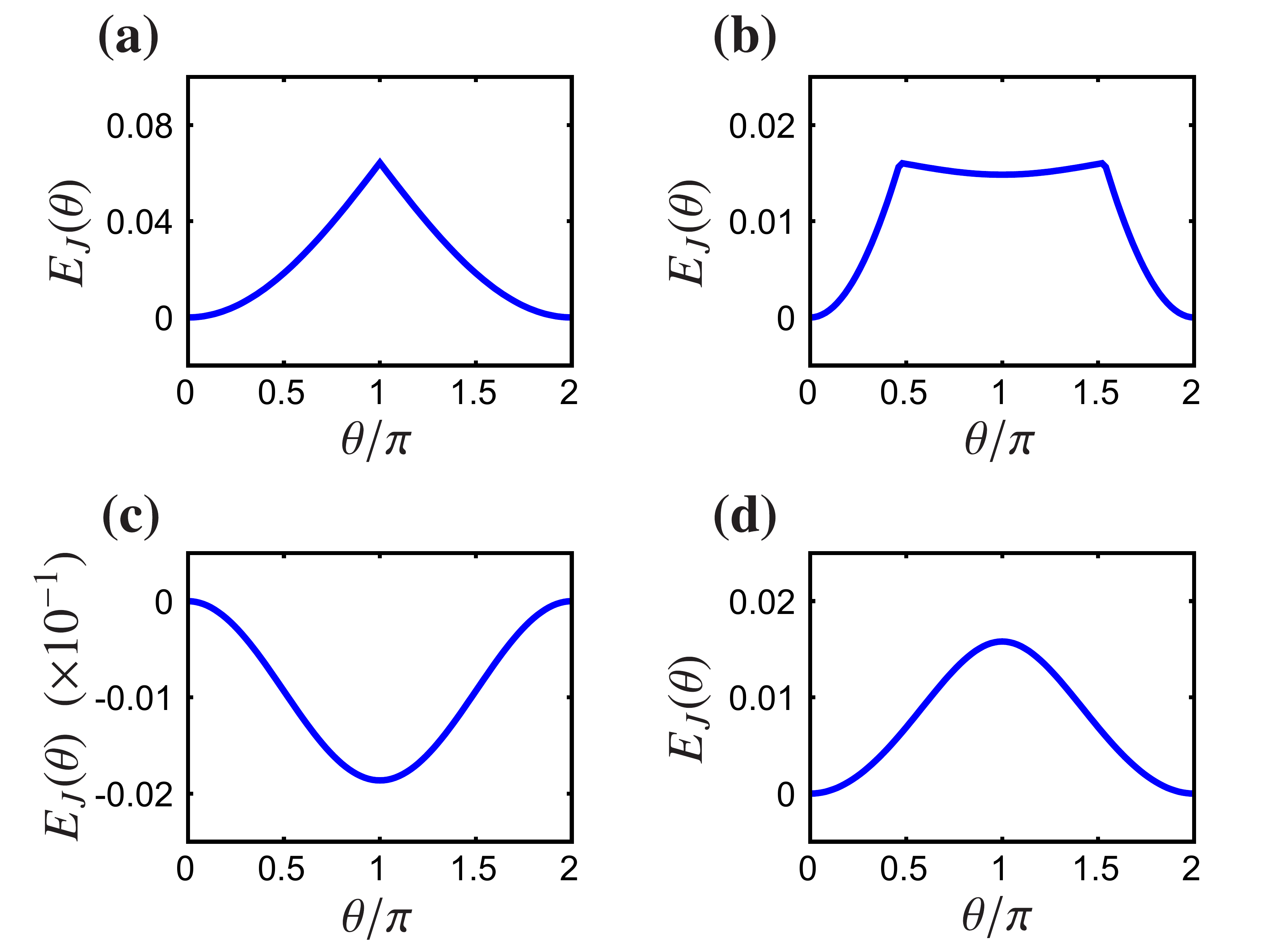}
    \caption{The Josephson energy-phase relation $E_J(\theta)$ of the ladder junction model when (a)~$t_\perp/t=0$,~~(b)~$t_\perp/t=0.1$,~~(c)~$t_\perp/t=0.3$,~~(d)~$t_\perp/t=1.0$. The cusp structure at $\theta = \pi$ in (a) is the consequence of 
    %derives from 
    the level crossing of two energy branches proportional to $\pm\cos\left( \theta/2 \right)$, 
    each of which has a $4\pi$-periodicity.
    %curves $\propto\pm\cos(\theta/2)$ which have 4$\pi$-periodicity.
    } \label{fig:josephson_energy}
  \end{center}
\end{figure}

First, let us look at the Josephson energy-phase relationship $E_{J}(\theta)$ of the system (\ref{eq:total}).
Figure \ref{fig:josephson_energy} shows $E_{J}(\theta)$ for several values of $t_\perp/t$, fixing the other parameters as $\Delta/t=0.5,~\mu/t=0.5$ and $\phi = \pi/3$. 
%In this calculation, we fix the number of sites as $N=100$ per chain. 
%From the calculation results, 
We find four qualitatively different types of behaviours of $E(\theta)$ as seen in Figs. \ref{fig:josephson_energy} (a)-(d).
%patterns of EPR, 
Figure \ref{fig:josephson_energy} (a) is exactly the topological Josephson energy-phase relationship with 4$\pi$-periodicity.
This is just because $t_{\perp}=0$; there are two independent Josephson junctions of the Kitaev chains.
% ($E_J(\theta)\propto\pm \cos(\theta/2)$), which unique to 1D topological ($p$-wave) superconductors, 
In Fig. \ref{fig:josephson_energy} (b), where $t_{\perp}/t = 0.1$, unusual behavior can be seen around $0.5 < \theta/\pi < 1.5$.
The Josephson energy $E_{J}(\theta)$ exhibits a transition from an increasing function to a decreasing function of $\theta$ at around $\theta \simeq 0.5/\pi$.
With further increasing $t_{\perp}/t$ to 0.3, $E_{J}(\theta)$ exhibits qualitatively different, but well-known $\theta$-dependence; the so-called $\pi$-junction as shown in Fig. \ref{fig:josephson_energy} (c).
Finally, if we take $t_{\perp}/t = 1.0$, the system switches back to the usual 0-junction property as shown in Fig. \ref{fig:josephson_energy} (d).

%mix of 4$\pi$-period and $\pi$-junction, (c) $\pi$-junction which means EPR has anomalous $\theta$ dependence like $E_J(\theta)\propto -\cos(\theta+\pi)$, and (d) 0-junction which represents normal EPR ($E_J(\theta)\propto -\cos(\theta)$).

\begin{figure}[htbp]
  \begin{center}
    \includegraphics[width=150mm]{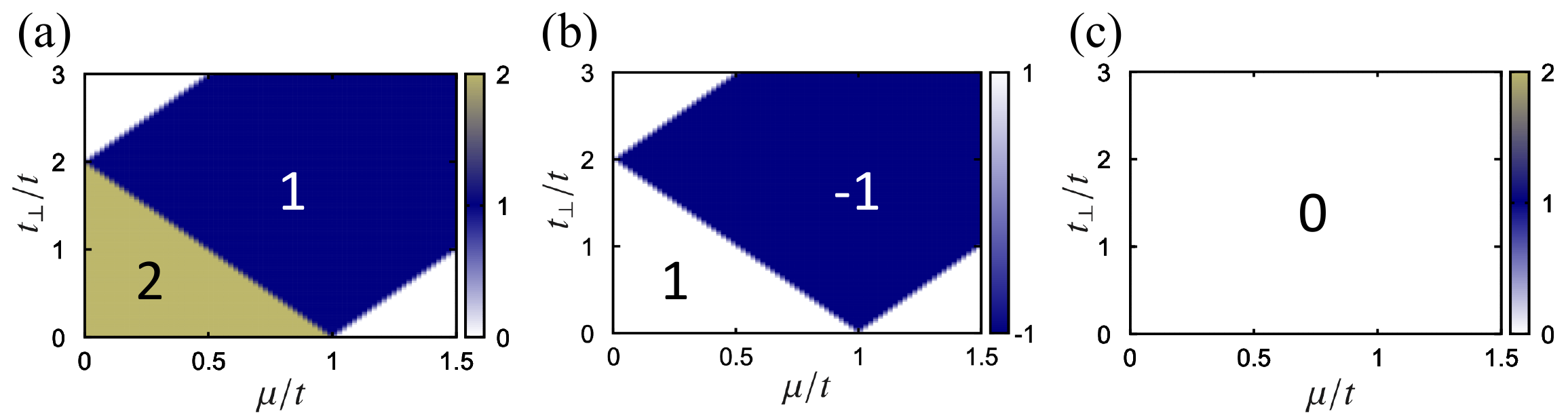}
    \caption{The topological phase diagrams of the Kitaev ladder system for (a)~$\phi = 0$,~~(b)~$\phi = \pi/3$,~~(c)~$\phi=\pi$, by using the Chern number $w$ for (a) and (c), and also the Majorana number ${\cal M}$ for (b). The integers in the figures indicate the Chern number in (a) and (c), and the Majorana number in (b). The system is in the topologically nontrivial phases when $w \neq 0$ or ${\cal M} = -1$.
    } \label{fig:topo_phase}
  \end{center}
\end{figure}

Here, the question arises; how the four characteristic behaviors found here depend on the model parameters, in particular $\mu$ and $t_{\perp}$.
Thus, we next look at the correlation between the types of $E_{J}(\theta)$ and the topological phases of the Kitaev ladder system.
Before that, however, it is necessary to recall that the phase diagram of the ladder system is modified when $\phi > 0$ because the symmetry class of the system changes from BDI to D due to the broken time-reversal and chiral symmetries, and changes back to BDI when $\phi=\pi$\cite{Ryu2010}.
Thus, for $0 < \phi < \pi$, the system is characterized by a $\mathbb{Z}_{2}$ invariant ${\mathcal M}$ called the Majorana number, instead of the Chern number $w$ for $\phi = 0$\cite{maiellaro2018}.
Following Maiellaro ${\it et~al.}$\cite{maiellaro2018}, we calculate the Chern number $w$ and the Majorana number ${\cal M}$ for several values of $\phi$ to see the effects of the interchain phase difference $\phi$ on the topological phase diagram of the Kitaev ladder system, and the results are summarized in Fig. \ref{fig:topo_phase}.
For $\phi = 0$, the well-known phase diagram of the Kitaev ladder system is reproduced as shown in Fig. \ref{fig:topo_phase} (a).
We also confirm that the Chern number $w$ is 0 for any $\mu$ and $t_{\perp}$ when $\phi = \pi$, thus the system is always in a topologically trivial phase, as shown in Fig. \ref{fig:topo_phase} (c).
If the symmetry class of the system is D when $0 < \phi < \pi$, the system is in a topologically nontrivial phase when ${\cal M} = -1$, %shown in Fig. \ref{fig:topo_phase} (b)
and the region where this is the case shown in Fig. \ref{fig:topo_phase} (b) coincides with that with $w=1$ when $\phi = 0$.
Now we can see the relation between the types of Josephson coupling and the topological phases realized in the Kitaev ladder system
based on the phase diagrams shown in Fig. \ref{fig:topo_phase}.

\begin{figure}[htbp]
  \begin{center}
    \includegraphics[height=100mm]{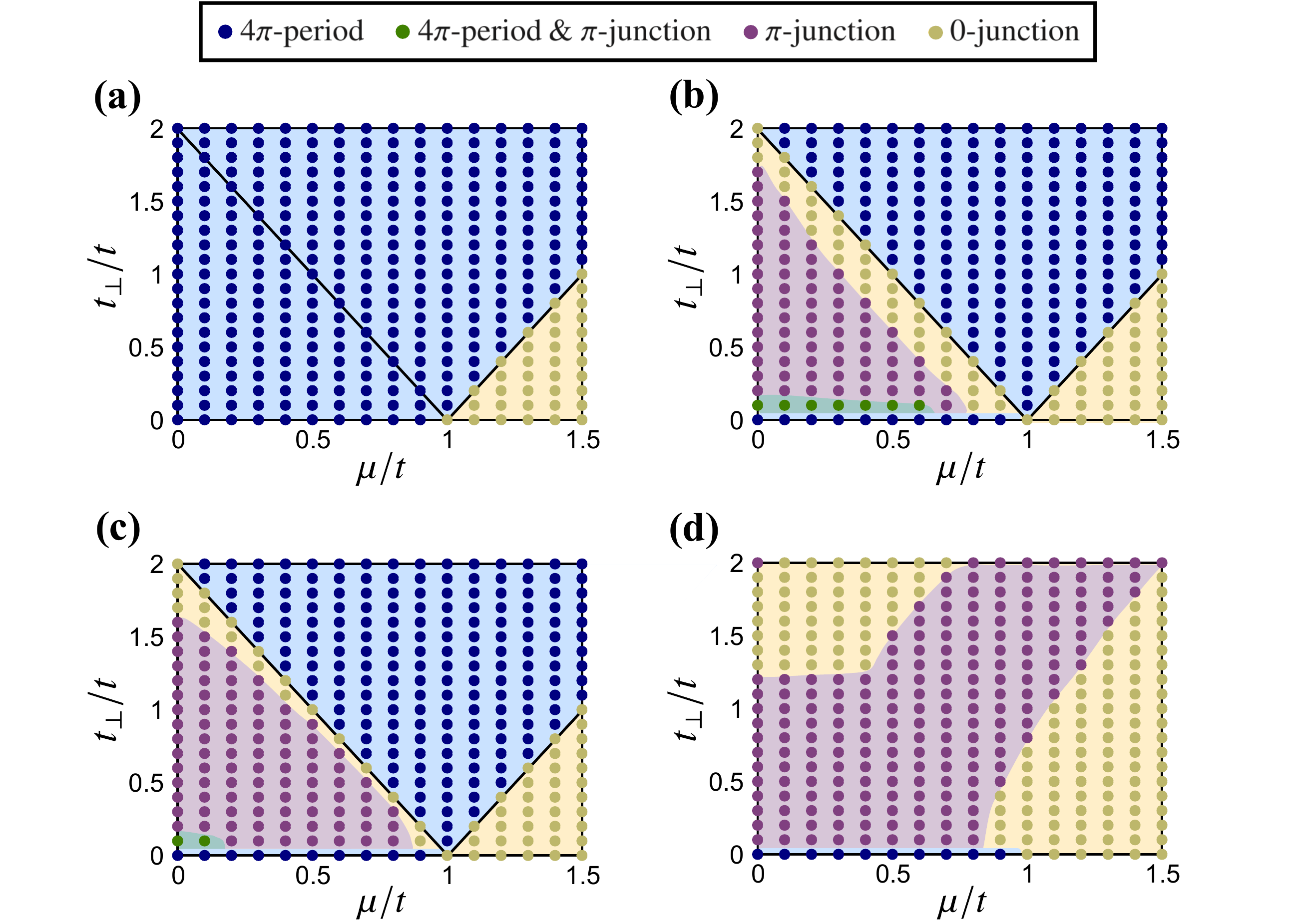}
    \caption{
    The correlation between the types of $E_{J}(\theta)$ on the phase diagram of the Kitaev ladder
    %$\mu,~t_\perp$ dependence of EPR when 
    for (a)~$\phi=0$,~~(b)~$\phi=\pi/3$,~~(c)~$\phi=2\pi/3$, and (d)~$\phi=\pi$. 
    Solid lines represent the boundary of the topological phase.}
    \label{fig:josephson_phase}
  \end{center}
\end{figure}

%the interchain phase difference $\phi$ changes the symmetry class of the ladder system
%
%that which pattern of EPR appear when we change the parameters $\mu,~t_\perp$ and $\phi$. 
Figure \ref{fig:josephson_phase} shows which type of $E_{J}(\theta)$ appears on the phase diagram of the Kitaev ladder system calculated for several values of $\phi$.
For $\phi = 0$ shown in Fig. \ref{fig:josephson_phase} (a), we can see that the region where the topological Josephson energy-phase relationship with 4$\pi$-periodicity (indicated by blue dots in the figure) appears coincides with the region with $w\neq 0$ in the phase diagram Fig. \ref{fig:topo_phase} (a).
In the remaining region in Fig. \ref{fig:josephson_phase} (a), where the $w=0$, the system exhibits the 0-junction property.
For $0 < \phi < \pi$ shown in Fig.  \ref{fig:josephson_phase} (b) and (c), the $\pi$-junction region appears (shown by red dots) in a fairly wide region of the phase diagram.
We notice that the region where the $\pi$-junction appears almost coincides with the region with the Majorana number is 1 where the system is in the topologically {\it{trivial}} state.
If we take $\phi = \pi$, the regions with the 4$\pi$-junction disappears because the the Chern number $w$ is 0, i.e., there is no MZMs, in the whole region of the phase diagram, as shown in Fig. \ref{fig:topo_phase} (c).
Instead, the $\pi$-junction region can be seen in wider range of the phase diagram.
We should note here that the unusual $E_{J}(\theta)$ found in Fig. \ref{fig:josephson_energy} is realized in the narrow region where $t_{\perp}/t$ is small shown in green dots in Fig. \ref{fig:josephson_phase} (b) and (c).
%a phase diagram of EPR, which represents parameter dependence of EPR pattern. Solid lines represent the boundary of the topological phase. 
%\phi=0の場合(a)には、Chern数が非ゼロの2つの領域で4\pi周期のEPRを、その他ではnormalのEPRを示すことがわかった。本文には載せていないが、Chern数、すなわちMajorana粒子の数に応じてEPRの振幅も変化する。0<phi<piの領域である(b),(c)では(a)で見られた4pi, normalに加えてpi-junctionと呼ばれるEPRを示し、t_\perpが微小な領域では4piとpiのmixが見られ、合計で4つものパターンが見られた。 (d)のphi=piではChern数が常にゼロであるため4pi周期のEPRは現れないが、その代わりに(b)(c)では見られなかった領域で広くpi-junctionが存在することが確認できる。また、muが小さくt_\perpが大きな領域にもpi-junctionが見られたことから、(b)(c)でもmu,t_\perpともに小さな領域だけでなく様々な領域に分布することが予想される。
%
From this result, it is found that 4$\pi$-period energy-phase relationship always appears in the region with $w\neq 0$, and $\pi$-junction appears in the region with $w=0$ or ${\cal M} = 1$.%, with relatively small $\mu/t$ and $t_\perp/t$.
%
%For the case of non-voltage biased ($\phi=0$), the period of EPR is 4$\pi$ due to the Majorana fermions at the edge of Kitaev ladders. On the other hand, for the case of voltage-biased ($\phi\neq0$),
%
%\newpage
%%%%%%%%%%%%%%%%%%%%%%%%%%%%%%%%%%%%%%%%%%%%%%
%\section{Discussion}
%%%%%%%%%%%%%%%%%%%%%%%%%%%%%%%%%%%%%%%%%%%%%%
%準粒子エネルギーとLDOS
%In order to investigate the origin of $\pi$-junction, we calculated energy spectra of Kitaev ladder for the parameter that $\Delta/t=1,~\mu/t=0.5$ and $\phi=\pi/3$, shown in figure \ref{fig:spectra}(a). In this figure, Majorana zero mode appear in $1<t_\perp/t<3$, where topological invariant is finite. On the other hand, in the case of $0<t_\perp/t<1$, finite but low-energy state appear. This state is found to be localized state at ladder edge from local DOS calculation in figure \ref{fig:spectra}(b). As mentioned in Nakosai \textit{et al}. \cite{nakosai2018}, two Majorana fermions at the edge of each chain deform a single Fermi-quasiparticle when $\phi\neq0$. Since these bound states are not seen in conventional superconducting wire, these states could cause $\pi$-junction and it is necessary to investigate more about these bound states.
More detailed studies on the origin of the various Josephson energy-phase relation $E_{J}(\theta)$ will be reported in a forthcoming paper.

%%%%%%%%%%%%%%%%%%%%%%%%%%%%%%%
%%%%%%%%%%%%%%%%%%%%%%%%%%%%%%%
%
%%%%%%%%%%%%%%%%%%%%%%%%%%%%%%%
%%%%%%%%%%%%%%%%%%%%%%%%%%%%%%%
%
%
%

%\begin{figure}[htbp]
%  \begin{center}
%    \includegraphics[height=55mm]{spectra_ldos.pdf}
%    \caption{(a) Perpendicular hopping dependence of energy spectra of the single Kitaev ladder when $\phi=\pi/3,~\Delta/t=\mu/t=0.5$, and the number of sites per chain is $N=30$. Red lines represent the spectra with the smallest and next smallest absolute value. (b) Local DOS (LDOS) at the edge of the Kitaev ladder, when $t_\perp/t=0.5$, $N=250$, site number is $i=1$, and other parameters are same as (a).} \label{fig:spectra}
%  \end{center}
%\end{figure}

%\begin{figure}[htbp]
%  \begin{center}
%    \includegraphics[height=120mm]{ldos_bound.pdf}
%    \caption{Local DOS at junction point of the ladder junction system for $\phi=0$, $\pi/3$ and $N=250$, where $4\pi$-junction and $\pi$-junction appear, respectively. Figures on the right side represent $\theta$ dependence of the energy for bound states at the edge of the junction system and the junction points. Other parameters such as chemical potential $\mu$ is fixed same as figure \ref{fig:spectra}.} \label{fig:bound}
%  \end{center}
%\end{figure}

%%%%%%%%%%%%%%%%%%%%%%%%%%%%%%%%%%%%%%%%%%%%%%
\section{Summary}
%%%%%%%%%%%%%%%%%%%%%%%%%%%%%%%%%%%%%%%%%%%%%%
In summary, we have investigated the Josephson energy-phase relationship in the Kitaev ladder systems by using numerical diagonalization, stimulated by the richer phase diagram and novel properties reported in the previous works \cite{wakatsuki2014,maiellaro2018,nakosai2018,Nehra2019}.
We have found the variety of Josephson energy-phase relationships, such as the topological $4\pi$-, $\pi$-, and conventional $0$-junctions. 
It is noteworthy that these Josephson current properties can be switched by externally controlling the interchain coupling $t_{\perp}$. 
This will open up new application possibilities \cite{demler1997,gingrich2016,Fukaya2020,baek2014} for the Kitaev superconducting chain other than topological quantum computation.
%We note here that these Josephson current characteristics can be switched by controlling the inter-chain coupling $t_{\perp}$.
%Thus it is expected to open up new application possibilities for the Kitaev superconducting chains.

%%%%%%%%%%%%%%%%%%%%%%%%%%%%%%%%%%%%%%%%%%%%%%
\section*{Acknowledgements}
%%%%%%%%%%%%%%%%%%%%%%%%%%%%%%%%%%%%%%%%%%%%%%
%This work was supported by JSPS %{\color{red}S}
%KAKENHI Grant Number JP21H01025.
%{\color{red} , 
%and also by JST CREST Grant Number JPMJCR18T2.
%ESICMM Grant Number JPMXP0112101004, 
%The work of P. N. was supported by project SOLID21, 
%and T. Y. was supported by JPS KAKENHI Grant Number JP18K04678.
%A. K., J. F., T. S., and H. T. acknowledge the support by the EU Horizon 2020 Program 
%under grant number 686056 (Novamag).
Some numerical computations were carried out at the Cyberscience Center, Tohoku University, Japan.
%%%%%%%%%%%%%%%%%%%%%%%%%%%%%%%%%%%%%%%%%%%%%%
% TODO: include funding information
\paragraph{Funding information}
This work was supported by JSPS %{\color{red}S}
KAKENHI Grant Number JP21H01025.
%Authors are required to provide funding information, including relevant agencies and grant numbers with linked author's initials. Correctly-provided data will be linked to funders listed in the \href{https://www.crossref.org/services/funder-registry/}{\sf Fundref registry}.
%%%%%%%%%%%%%%%%%%%%%%%%%%%%%%%%%%%%%%%%%%%%%%

% TODO:
% Provide your bibliography here. You have two options:

% FIRST OPTION - write your entries here directly, following the example below, including Author(s), Title, Journal Ref. with year in parentheses at the end, followed by the DOI number.
%\begin{thebibliography}{99}
%\bibitem{1931_Bethe_ZP_71} H. A. Bethe, {\it Zur Theorie der Metalle. i. Eigenwerte und Eigenfunktionen der linearen Atomkette}, Zeit. f{\"u}r Phys. {\bf 71}, 205 (1931), \doi{10.1007\%2FBF01341708}.
%\bibitem{arXiv:1108.2700} P. Ginsparg, {\it It was twenty years ago today... }, \url{http://arxiv.org/abs/1108.2700}.
%\end{thebibliography}

% \bibliographystyle{SciPost_bibstyle} % Include this style file here only if you are not using our template
\bibliography{Kanehira.bib}

\nolinenumbers

\end{document}